\documentclass[10pt,conference]{IEEEtran}
\usepackage{epsfig,setspace,amsmath,epsf,amssymb,bm,theorem,cite, graphicx, epstopdf, algorithm, algpseudocode,float,color}
\usepackage{multirow}
\usepackage{authblk}
\usepackage[table,xcdraw]{xcolor}
\usepackage{mathtools}
\usepackage{caption}
\usepackage{subcaption}
\usepackage{float}

\newtheorem{theorem}{Theorem}

\newtheorem{remark}{Remark}
\newtheorem{lemma}{Lemma}

\newenvironment{Proof}[1]{\medskip\par\noindent{\bf Proof:\,}\,#1}{{\mbox{\,$\blacksquare$}\par}}

\newcommand{\bz}{{\mathbf{Z}}}
\newcommand{\bw}{{\mathbf{W}}}
\newcommand{\bQ}{{\mathbf{Q}}}
\newcommand{\bS}{{\mathbf{S}}}
\newcommand{\be}{{\mathbf{e}}}
\newcommand{\bD}{{\mathbf{\Delta}}}

\newcommand{\bi}{{\mathbf{I}}}
\newcommand{\bp}{{\mathbf{P}}}

\IEEEoverridecommandlockouts

\newcommand{\st}{{\text{s.t.}}}

\allowdisplaybreaks

\begin{document}

\title{Private Read Update Write (PRUW) with \\ Storage Constrained Databases\thanks{This work was supported by ARO Grant W911NF2010142, and NSF Grants CCF 17-13977 and ECCS 18-07348.}}

\author{Sajani Vithana \qquad Sennur Ulukus\\
\normalsize Department of Electrical and Computer Engineering\\
\normalsize University of Maryland, College Park, MD 20742\\
\normalsize  \emph{spallego@umd.edu} \qquad \emph{ulukus@umd.edu}}

\maketitle

\begin{abstract}
We investigate the problem of private read update write (PRUW) in relation to federated submodel learning (FSL) with storage constrained databases. In PRUW, a user privately reads a submodel from a system of $N$ databases containing $M$ submodels, updates it locally, and writes the update back to the databases without revealing the submodel index or the value of the update. The databases considered in this problem are only allowed to store a given amount of information specified by an arbitrary storage constraint. We provide a storage mechanism that determines the contents of each database prior to the application of the PRUW scheme, such that the total communication cost is minimized. We show that the proposed storage scheme achieves a lower total cost compared to what is achieved by using \emph{coded storage} or \emph{divided storage} to meet the given storage constraint. 
\end{abstract}

\section{Introduction}

Federated submodel learning (FSL) \cite{billion, paper1, secureFSL, rw_jafar, ourICC, dropout, aggregation} is a form of federated learning (FL)  \cite{FL1, Advances, FL2,magazine, keyboard} where the model is divided into multiple submodels so that each user is able to download and update only the specific submodel(s) that can be trained by the user's local data. FSL consists of two phases in communication, namely, the reading phase in which the users download the required submodel, and the writing phase in which the users upload the generated update. Original FL which requires the users to download and update the entire model is inefficient compared to FSL in cases where the users do not have the types of data suitable to train the entire model. Although FSL is efficient in terms of communication cost and processing power of local users, it introduces an important issue with respect to user privacy. The submodel that a given user updates may leak information on the type of data the user has. Moreover, the values of the updates uploaded by a user may leak information about the local data of the user, as in FL \cite{recent, comprehensive, featureLeakage, InvertingGradients, client, DPFL, language, PracticalSecureAgg, avgDP,cpSGD}. Consequently, in order to guarantee the privacy of a given user, the index of the updating submodel as well as the values of the updates must be kept private from databases. The combined process of privately reading, updating and writing is known as private read update write (PRUW). 

Existing works \cite{billion, paper1, rw_jafar, dropout,ourICC,aggregation} provide PRUW schemes with different notions of privacy\cite{DP, OTP}. PRUW with information-theoretic privacy is equivalent to the problem of private information retrieval (PIR) in the reading phase, see e.g., \cite{original, PIR, coded,  ravi_storage_constrained, utah, ITW_paper, Kumar_PIRarbCoded, YamamotoPIR, VardyConf2015, MultiroundPIR, heteroPIR, decentralized, chaoTian, leaky, SPIR, PSI, colluding, CodeColludeByzantinePIR, sideinfo, MMPIR, MMPIR_PrivateSideInfo, evesdroppers, byzantine, semanticPIR, singleDB,XSTPIR, asympXSTP, smallfields, SecureStorage}. The PRUW scheme that is based on information-theoretic privacy, with the lowest known total cost (reading+writing) so far, is presented in \cite{dropout} and \cite{ourICC}. This scheme requires $N$ databases that store $ML$ bits where $M$ is the number of submodels and $L$ is the size of a submodel. However, in practice, the available databases may not have the capacity to store $ML$ bits since the model sizes can be large in general. Thus, the existing schemes cannot be directly defined on databases with given storage limitations. In this work, we investigate efficient storage mechanisms that can be used for PRUW on databases with given storage constraints. 

The works most closely related to ours are \cite{rw_jafar, ourICC, dropout, ravi_storage_constrained, utah,ITW_paper,coded}. FSL with coded/uncoded storage is studied in \cite{rw_jafar, ourICC, dropout} under general privacy and security constraints. For storage constrained PIR, \cite{coded} presents a PIR scheme with \emph{coded storage}, while \cite{ravi_storage_constrained, utah} provide uncoded schemes with \emph{divided storage}. A scheme that utilizes \emph{both coded and divided} storage in order to minimize the storage cost is presented in \cite{ITW_paper}.

In this paper, we investigate the problem of PRUW with homogeneous storage constrained databases, where the common storage capacity of each database is specified as a fraction $\mu$ of $ML$. PRUW with storage constrained databases consists of two main stages: 1) determining the contents of each database, 2) employing a PRUW scheme. In this work, we focus on the first phase of determining the storage of each database such that the optimized version of the PRUW scheme in \cite{dropout} can be applied in the next stage while satisfying the given storage constraint. The storage mechanism we propose takes the given storage constraint $\mu$ as an input and determines the contents of each database prior to the read/write process, such that the total communication cost (reading+writing) is minimized. 

A given storage constraint can be met by \emph{dividing} the submodels according to \cite{ravi_storage_constrained, utah} and storing subsets of them in each database, or by utilizing \emph{coded} storage \cite{dropout}. In this work, we provide a \emph{hybrid} storage mechanism that combines both coding and dividing. We show that the hybrid mechanism achieves a lower total cost compared to each of the two individual mechanisms for any given storage constraint $\mu$.
We also provide lower bounds on the achievable total costs of each individual storage mechanism, and illustrate how the proposed hybrid scheme achieves a lower total cost.

\section{Problem Formulation}

We consider a FL model consisting of $M$ independent submodels, each containing $L$ bits, stored in a system of $N$, $N\geq4$, non-colluding databases. Each database has a storage capacity of $\mu ML$ bits, where $\mu\in\left[\frac{1}{N-3},1\right]$. Each database must satisfy $H(S_n)\leq \mu ML$, where $S_n$ is the content of database $n$, $n\in\{1,\dotsc,N\}$. At any given time instance, a single user reads, updates and writes a single submodel of interest, while keeping the submodel index and the value of the update private from all databases and other users. 

\emph{Privacy of the submodel index:} No information on the index of the submodel being updated $\theta$ is allowed to leak to any of the databases, i.e., for each $n$,
\begin{align}
    I(\theta^{[t]};Q_n^{[t]},U_n^{[t]}|Q_n^{[1:t-1]},S_n^{[1:t-1]},U_n^{[1:t-1]})=0,
\end{align}
where $Q_n^{[t]}$ and $U_n^{[t]}$ are the query and update sent by the user to database $n$ at time $t$ in the reading and writing phases.

\emph{Privacy of the value of the update:} No information on the value of the update is allowed to leak to any of the databases, i.e., for each $n$,
\begin{align}
    I(\Delta_{\theta}^{[t]};U_n^{[t]}|Q_n^{[1:t]},S_n^{[1:t-1]},U_n^{[1:t-1]})=0,
\end{align}
where $\Delta_{\theta}$ is the update generated by the user.

\emph{Security of submodels:} No information on the submodels is allowed to leak to any of the databases, i.e., for each $n$,
\begin{align}
    I(\bw_{1:M}^{[t]};S_n^{[t]})=0,
\end{align}
where $\bw_k^{[t]}$ is the $k$th submodel at time $t$.

\emph{Correctness in the reading phase:} The user should be able to correctly decode the required submodel from the answers received in the reading phase, i.e., 
\begin{align}
H(\bw_{\theta}^{[t-1]}|Q_{1:N}^{[t]},A_{1:N}^{[t]})=0,
\end{align}
where $A_n^{[t]}$ is the answer from database $n$ at time $t$.

\emph{Correctness in the writing phase:} Each parameter $i$, $i\in\{1,\dotsc,L\}$ of $\bw_{\theta}$ must be correctly updated at time $t$ as,
\begin{align}
    \bw_{\theta,i}^{[t]}=\bw_{\theta,i}^{[t-1]}+\Delta_{\theta,i}.
\end{align}

The reading and writing costs are defined as $C_R=\frac{\mathcal{D}}{L}$ and $C_W=\frac{\mathcal{U}}{L}$, respectively, where $\mathcal{D}$ is the total number of bits downloaded in the reading phase and $\mathcal{U}$ is the total number of bits uploaded in the writing phase. The total cost $C_T$ is the sum of the reading and writing costs, i.e., $C_T=C_R+C_W$.

\section{Main Result}

In this section, we present the achievable total cost of the proposed storage mechanism. Let $C_T(\mu)$ be the total cost corresponding to the storage constraint $\mu$.

\begin{theorem}\label{thm1}
For any given $N$ and $\mu\in\left[\frac{1}{N-3},1\right]$, the achievable total cost of the proposed scheme is given by the boundary of the lower convex hull of $(\mu,C_T(\mu))$ pairs given by,
\begin{align}
    &\left(\mu=\frac{r}{NK_r}, C_T(\mu)=\frac{4r}{r-K_r-1}\right),\quad r=4,\dotsc,N,\nonumber\\ 
    &\quad K_r=1,\dotsc,r-3, \ \st \ (r-K_r-1)\!\!\!\mod2=0.
\end{align}
\end{theorem}

\begin{remark}
The two main constrained storage mechanisms (divided storage and coded storage) are subsets of the proposed hybrid storage mechanism. Divided storage corresponds to cases where $r$ is even in $\{4,\dotsc,N\}$ and $K_r=1$, while coded storage corresponds to $r=N$ and $K_r\in\{1,\dotsc,N-3\}$, with $(N-K_r-1)\!\!\mod2=0$. Since the achievable total cost of the proposed scheme is characterized by the boundary of the lower convex hull of all achievable points, the total cost of the proposed scheme is less than or equal to that of the two main storage mechanisms for each $\mu$. 
\end{remark}

\section{Proposed Storage Mechanism}

In this section, we present the proposed storage mechanism that specifies the specific storage in each database according to the given storage constraint $\mu$, prior to the application of the PRUW scheme. First, we present the following lemma, which is a crucial component of the proposed scheme.

\begin{lemma}\label{lem1}
Let $(\mu_1,C_T(\mu_1))$ and $(\mu_2,C_T(\mu_2))$ be two pairs of storage constraints and corresponding achievable total costs. Then, the pair $(\mu,C_T(\mu))$ is also achievable where
\begin{align}
    \mu&=\gamma\mu_1+(1-\gamma)\mu_2\\
    C_T(\mu)&=\gamma C_T(\mu_1)+(1-\gamma)C_T(\mu_2) 
\end{align}
for any $\gamma\in[0,1]$.
\end{lemma}

\begin{Proof}
Since $(\mu_1,C_T(\mu_1))$ and $(\mu_2,C_T(\mu_2))$ are achievable, let $S_1$ and $S_2$ be the schemes that produce the achievable pairs $(\mu_1,C_T(\mu_1))$ and $(\mu_2,C_T(\mu_2))$, respectively. A new scheme can be generated by applying $S_1$ on a $\gamma$ fraction of bits of all submodels and $S_2$ on the rest of the bits. The storage of this scheme is, $\gamma ML\mu_1+(1-\gamma)ML\mu_2=\mu ML$ bits. The corresponding total cost of the combined scheme is 
\begin{align}
    C_T\!\!=\!\frac{\gamma LC_T(\mu_1)\!+\!(\!1\!-\!\gamma)LC_T(\mu_2)}{L}\!\!=\!\gamma C_T(\mu_1)\!+\!(\!1\!-\!\gamma)C_T(\mu_2),
\end{align}
completing the proof.
\end{Proof}

Next, we need to obtain the optimized version of the PRUW scheme in \cite{dropout} in order to determine the contents of each database, on which the scheme is applied.

\subsection{Optimized PRUW Scheme}\label{scheme}

The contents of a single subpacket in database $n$, $n\in\{1,\dotsc,r\}$ of the scheme in \cite{dropout} with a subpacketization of $y$ and $x+1$ noise terms with no dropouts is given by
\begin{align}\bS_n=
    \begin{bmatrix}
    \sum_{i=1}^K\frac{1}{f_{1,i}-\alpha_n}\begin{bmatrix}
    \bw_{1,1}^{[i]}\\\vdots\\\bw_{M,1}^{[i]}
    \end{bmatrix}+\sum_{j=0}^x\alpha_n^j\bi_{1,j}\\
    \vdots\\
    \sum_{i=1}^K\frac{1}{f_{y,i}-\alpha_n}\begin{bmatrix}
    \bw_{1,y}^{[i]}\\\vdots\\\bw_{M,y}^{[i]}
    \end{bmatrix}+\sum_{j=0}^x\alpha_n^j\bi_{y,j}
    \end{bmatrix},\label{storage}
\end{align}
with random noise vectors $\bi_{i,j}$ (for security of submodels \cite{OTP}) and distinct constants $f_{i,j}, \alpha_n$. In the reading phase, the user sends queries $\bQ_{n,\ell}$, $\ell\in\{1,\dotsc,K\}$ to retrieve each of $\bw^{[1]}_{[\theta],1},\dotsc,\bw^{[K]}_{[\theta],y}$, where $\theta$ is the required submodel index, 
\begin{align}
    \bQ_{n,\ell}=\begin{bmatrix}\frac{\prod_{i=1,i\neq \ell}^{K}(f_{1,i}-\alpha_n)}{\prod_{i=1,i\neq \ell}^{K}(f_{1,i}-f_{1,\ell})}\be_M(\theta)+\prod_{i=1}^K (f_{1,i}-\alpha_n)\bz_{1,\ell}\\
    \vdots\\
    \frac{\prod_{i=1,i\neq \ell}^{K}(f_{y,i}-\alpha_n)}{\prod_{i=1,i\neq \ell}^{K}(f_{y,i}-f_{y,\ell})}\be_M(\theta)+\prod_{i=1}^K (f_{y,i}-\alpha_n)\bz_{y,\ell}
\end{bmatrix},\label{query}
\end{align}
where $\bz_{i,j}$ are random noise vectors. The databases send the answers $A_{n,\ell}$, $\ell\in\{1,\dotsc,K\}$ given by,
\begin{align}\label{ans}
    A_{n,\ell}&=\bS_n^T\bQ_{n,\ell}=\sum_{i=1}^y\frac{1}{f_{i,\ell}-\alpha_n}\bw_{\theta,i}^{[\ell]}+\sum_{j=0}^{K+x}\alpha_n^j\Tilde{\bi}_j, 
\end{align}
where $\Tilde{\bi}_i$ are combinations of random noise terms. Using the answers of all $r$ databases, $\{\bw^{[\ell]}_{\theta,1},\dotsc,\bw^{[\ell]}_{\theta,y}\}$ can be obtained for each $\ell$ if $r=y+x+K+1$. The resulting reading cost is 
\begin{align}
C_R=\frac{Kr}{Ky}=\frac{r}{r-x-K-1}.
\end{align}

In the writing phase, the user sends $K$ bits to each of the $r$ databases, which are linear combinations of $y$ update bits,
\begin{align}\label{update}
    U_{n,\ell}=\sum_{j=1}^y\prod_{i=1,i\neq j}^y(f_{i,\ell}-\alpha_n)\Tilde{\Delta}_{j,\ell}^{[\theta]}+\prod_{i=1}^y(f_{i,\ell}-\alpha_n)\hat{z}_{\ell},
\end{align}
where $\Tilde{\bD}_{j,\ell}^{[\theta]}=\frac{\prod_{i=1,i\neq \ell}^K (f_{j,i}-f_{j,\ell})}{\prod_{i=1,i\neq j}^y (f_{i,\ell}-f_{j,\ell})}\bD_{j,\ell}^{[\theta]}$ for $j\in\{1,\dotsc,y\}$. Once database $n$ receives the update bits, it calculates the incremental update with the aid of the two matrices given by,
\begin{align}
    \Omega_{n,\ell}\!&=\!\text{diag}\!\left(\!\frac{\prod_{r\in\mathcal{F}} (\alpha_r\!-\!\alpha_n)}{\prod_{r\in\mathcal{F}} (\alpha_r\!-\!f_{1,\ell})}\!\mathbf{1}_M,\dotsc,\frac{\prod_{r\in\mathcal{F}} (\alpha_r\!-\!\alpha_n)}{\prod_{r\in\mathcal{F}} (\alpha_r\!-\!f_{y,\ell})}\!\mathbf{1}_M\!\right)\\
    \Tilde{\bQ}_{n,\ell}\!&=\!\text{diag}\!\left(\!\frac{1}{\prod_{i=1}^K(f_{1,i}\!-\!\alpha_n)}\!\mathbf{1}_M,\dotsc,\frac{1}{\prod_{i=1}^K(f_{y,i}\!-\!\alpha_n)}\!\mathbf{1}_M\!\right)\nonumber\\
    &\qquad\qquad\qquad\qquad\qquad\times\bQ_{n,\ell},
\end{align}
where $\Omega_{n,\ell}$ is the null shaper in \cite{dropout} with $|\mathcal{F}|=x-y$ and $\Tilde{\bQ}_{n,\ell}$ is the scaled query vector. $\mathbf{1}_M$ is the vector of all ones of size $1\times M$. The incremental update is calculated as,
\begin{align}
    \bar{U}_{n,\ell}&=\Omega_n\times U_{n,\ell}\times \Tilde{\bQ}_{n,\ell}\\
    &=\begin{bmatrix}
        \frac{1}{f_{1,\ell}-\alpha_n}\bD_{1,\ell}^{[\theta]}\be_M(\theta)+\bp_{\alpha_n}^{[1]}(x)\\
        \vdots\\
        \frac{1}{f_{y,\ell}-\alpha_n}\bD_{y,\ell}^{[\theta]}\be_M(\theta)+\bp_{\alpha_n}^{[y]}(x)
    \end{bmatrix},
\end{align}
where $P^{[j]}_{\alpha_n}(i)$ is a polynomial of $\alpha_n$ of degree $i$, indexed by $j$. Then, the submodels are updated by $\bS_n(t)=\bS_n(t-1)+\sum_{\ell=1}^K \bar{U}_{n,\ell}$. The resulting writing and total costs are,
\begin{align}
    C_W&=\frac{K(r-(x-y))}{Ky}=\frac{2r-2x-K-1}{r-x-K-1},\\
    C_T&=C_R+C_W=\frac{3r-2x-K-1}{r-x-K-1}.
\end{align}

Note that the total cost is an increasing function of $x$ since $\frac{dC_T}{dx}=\frac{r+K+1}{(r-x-K-1)^2}>0$. Since $x\geq y$ must be satisfied by $x$ in order to write to $y$ parameters using a single bit, the optimum value of $x$ that minimizes the total cost is,
\begin{align}\label{sub_noise}
    x=\begin{cases}
        y=\frac{r-K-1}{2}, \quad & \text{if $r-K-1$ is even},\\
        y+1=\frac{r-K}{2}, \quad & \text{if $r-K-1$ is odd}.
\end{cases}
\end{align}
The resulting total costs of the two cases are,
\begin{align}\label{totalcost2}
    C_T=\begin{cases}
    \frac{4r}{r-K-1}, \quad & \text{if $r-K-1$ is even},\\
    \frac{4r-2}{r-K-2}, \quad & \text{if $r-K-1$ is odd}.
    \end{cases}
\end{align}
Note that since the subpacketization $y\geq1$, $r$ and $K$ must satisfy,
\begin{align}\label{range}
    1\leq K\leq\begin{cases}
    r-3, \quad & \text{if $r-K-1$ is even},\\
    r-4, \quad & \text{if $r-K-1$ is odd}.
    \end{cases}
\end{align}

\subsection{Proposed Storage Mechanism}\label{proposed}

For a given $N$ we first find the basic achievable pairs of $(\mu,C_T(\mu))$ as follows. Let $\mu=\frac{r}{NK_r}$ for $r=4,\dotsc,N$ and $K_r=1,\dotsc,r-3$. For a given $\mu$ with a given $r$ and $K_r$, following steps need to be followed in order to perform PRUW while meeting the storage constraint: 
\begin{enumerate}
    \item Divide the $L$ bits of each submodel into $N$ sections and label them as $\{1,\dotsc,N\}$.
    \item Allocate sections $n:(n-1+r)\!\!\mod N$ to database $n$ for $n\in\{1,\dotsc,N\}$.\footnote{The indices here follow a cyclic pattern, i.e., if $(n-1+r)\!\!\mod N<n$, $n:(n-1+r)\!\!\mod N$ implies $\{n,\dotsc,N,1,\dotsc,(n-1+r)\!\!\mod N\}$.}
    \item Use the storage specified in \eqref{storage} with $K=K_r$ and $x,y$ given in \eqref{sub_noise} to encode each of the allocated sections of all submodels. Note that a given coded bit of a given section of each submodel stored across different databases contains the same noise polynomial that only differs in $\alpha_n$.
    \item Use the PRUW scheme described in Section~\ref{scheme} on each of the subsets of $n:(n-1+r)\!\!\mod N$ databases to read/write to section $(n-1+r)\mod N$ of the required submodel for $n\in\{1,\dotsc,N\}$.
\end{enumerate}

For each $\mu=\frac{r}{NK_r}$, $r=4,\dotsc,N$, $K_r=1,\dotsc,r-3$, the above process gives an achievable $(\mu,C_T(\mu))$ pair, where $C_T(\mu)$ is given as follows using \eqref{totalcost2},
\begin{align}\label{Tmu}
    C_T(\mu)=\begin{cases}
    \frac{4r}{r-K_r-1}, \quad & \text{if $r-K_r-1$ is even},\\
    \frac{4r-2}{r-K_r-2}, \quad & \text{if $r-K_r-1$ is odd}.
    \end{cases}
\end{align}
Note that the above two cases,  which correspond to the value of $(r-K_r-1)\!\!\mod2$, are a result of two different schemes. The case with even values of $r-K_r-1$ has a subpacketization that is equal to the degree of noise polynomial in storage, which does not require the null shaper, while the case with odd values of $r-K_r-1$ contains two more noise terms than the subpacketization, which requires the null shaper; see \eqref{sub_noise}. The scheme corresponding to odd values of $r-K_r-1$ is inefficient compared to the even case due to the additional noise term present in storage. This observation combined with Lemma~\ref{lem1} results in the following lemma.

\begin{lemma}\label{lem2}
For a given $\mu=\frac{r}{NK_r}$, if $r$ and $K_r$ are such that $r-K_r-1$ is odd, it is more efficient to perform a linear combination of two PRUW schemes with nearest two even $r^{[i]}-K_r^{[i]}-1$, $i=1,2$, instead of performing direct PRUW with the given $r$ and $K_r$, while satisfying the same storage constraint $\mu$, i.e., with $\mu_1=\frac{r^{[1]}}{NK_r^{[1]}}$ and $\mu_2=\frac{r^{[2]}}{NK_r^{[2]}}$.  
\end{lemma}

\begin{Proof}
For a given $\mu=\frac{r}{NK_r}$, the nearest $\mu_1=\frac{r^{[1]}}{NK_r^{[1]}}$ is $\frac{r-1}{NK_r}$, since \eqref{range} with $K$ replaced by $K_r$ needs to be satisfied for the PRUW scheme to work. Similarly, $\mu_2=\frac{r+1}{NK_r}$. Let $C_T(\mu)$, $C_T(\mu_1)$ and $C_T(\mu_2)$ be the total costs incurred by the scheme with $\mu$, $\mu_1$ and $\mu_2$, respectively. From \eqref{Tmu}, we have,
\begin{align}
    C_T(\mu)\!=\!\frac{4r\!-\!2}{r\!\!-\!K_r\!-\!2}\!\!>\!C_T(\mu_1)\!=\!\frac{4r\!-\!4}{r\!\!-\!K_r\!-\!2}\!\!>\!C_T(\mu_2)\!=\!\frac{4(r\!+\!1)}{r\!\!-\!K_r}.
\end{align}
Note that $\mu_1<\mu<\mu_2$. From Lemma~\ref{lem1}, there exists some $\gamma\in[0,1]$ that allocates the storage for the two PRUW schemes corresponding to $\mu_1$ and $\mu_2$ that achieves the same storage constraint as $\mu$, and results in a total cost of $\gamma C_T(\mu_1)+(1-\gamma)C_T(\mu_2)$, that satisfies,
\begin{align}
    C_T(\mu_2)<\gamma C_T(\mu_1)+(1-\gamma)C_T(\mu_2)<C_T(\mu_1)<C_T(\mu),
\end{align}
completing the proof. 
\end{Proof}

Once the basic $(\mu,C_T(\mu))$ pairs corresponding to $\mu=\frac{r}{NK_r}$ for $r=4,\dotsc,N$, $K_r=1,\dotsc,r-3$ with $(r-K_r-1)\mod2=0$ are obtained, the achievable total cost of the proposed scheme for any $\mu$ is characterized by the boundary of the lower convex hull of the above basic $(\mu,C_T(\mu))$ pairs, using Lemma~\ref{lem1}, denoted by $T_{ach}$. Therefore, for a given $N$ and $\mu$, the proposed PRUW storage mechanism is obtained by utilizing the correct linear combination of PRUW schemes that correspond to the nearest two basic $(\mu,C_T(\mu))$ pairs on $T_{ach}$. The resulting total cost is $T_{ach}(\mu)$. 

\begin{figure}[t]
    \centering
    \includegraphics[scale=0.63]{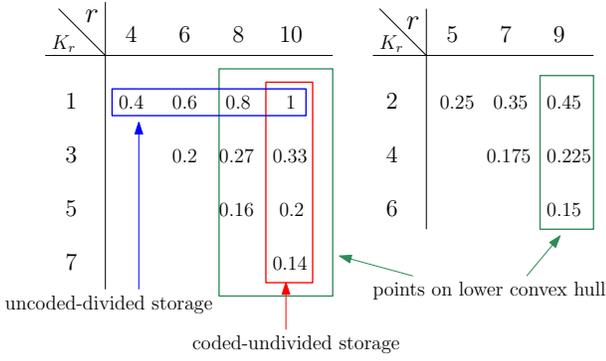}
    \caption{All possible pairs of $(r,K_r)$ and corresponding values of $\mu$ for $N=10$.}
    \label{fig2}
    \vspace*{-0.5cm}
\end{figure}

\begin{figure}[t]
    \centering
    \includegraphics[scale=0.63]{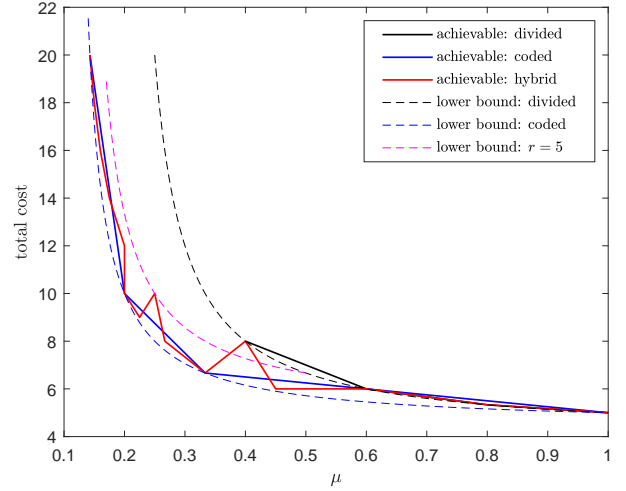}
    \caption{Achievable total costs and lower bounds of divided, coded and hybrid schemes for $N=10$, before obtaining the convex hull boundary.}
    \label{fig3}
    \vspace*{-0.5cm}
\end{figure}

\begin{figure}[t]
    \centering
    \includegraphics[scale=0.63]{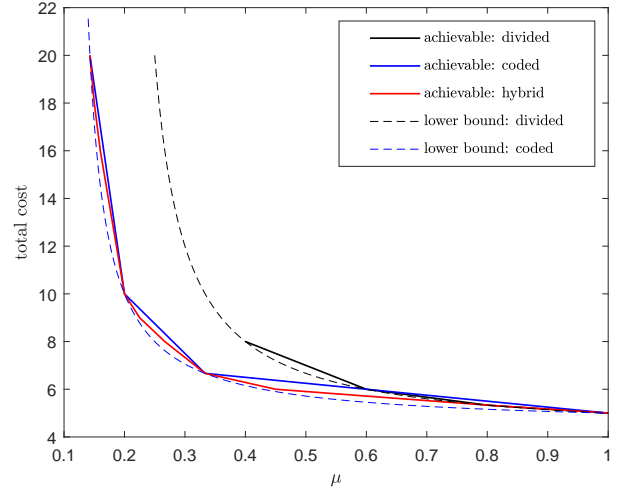}
    \caption{Lowest achievable costs of coded, divided and hybrid schemes for $N=10$.}
    \label{fig4}
    \vspace*{-0.5cm}
\end{figure}

\section{Lower Bounds on Achievable Costs}

In this section, we provide lower bounds on the achievable costs derived in Section~\ref{proposed}. Based on Lemma~\ref{lem2} and \eqref{Tmu}, for a given $K_r$, the achievable total cost of the proposed scheme for a given $N$ and $\mu\leq \frac{1}{K_r}$ can be lower bounded as,
\begin{align}\label{lbcoded}
    C_T(\mu,K_r)\!\geq\!\frac{4r}{r-K_r-1}\!=\!\frac{4N\mu}{N\mu-1-\frac{1}{K_r}}\!=\!LB(\mu,K_r),
\end{align}
since $LB(\mu,K_r)$ is a convex function of $\mu$ and the achievable total cost is a piecewise linear function with points corresponding to $\mu=\frac{r}{NK_r}$ with $(r-K_r-1)\mod2=0$ on $LB(\mu,K_r)$. Similarly, for a given $r$, the achievable total cost of any $\mu\leq \frac{r}{N}$ is lower bounded as,
\begin{align}\label{lbuncoded}
    C_T(\mu,r)\!\geq\!\frac{4r}{r-K_r-1}\!=\!\frac{4N\mu}{N\mu-1-\frac{N\mu}{r}}\!=\!LB(\mu,r).
\end{align}

Note that the two storage mechanisms defined by divided storage ($K_r=1$, $r<N$) and coded storage ($K_r>1$, $r=N$) are subsets of the proposed storage mechanism where $K_r\geq1$ and $r\leq N$. The total cost of the divided storage mechanism is lower bounded by $LB_d(\mu)=\frac{4N\mu}{N\mu-2}$, while that of coded storage is lower bounded by $LB_c(\mu)=\frac{4N\mu}{N\mu-1-\mu}$. Clearly, the lower bound of the coded scheme is less than that of the divided storage scheme except at $\mu=1$, where the two bounds are the same. For all other cases ($K_r>1$, $r<N$), the lower bounds $LB(\mu,r)$ and $LB(\mu,K_r)$ satisfy,
\begin{align}
LB_c(\mu)\leq LB(\mu,K_r), \quad LB(\mu,r)\leq LB_d(\mu),    
\end{align}
For each $\mu\in\left[\frac{1}{N-3},1\right]$: Even though the coded scheme is better in terms of the lower bounds, the achievable costs show that the divided storage scheme performs better at larger values of $\mu$, as shown in Fig.~\ref{fig4} for $N=10$. For the same example with $N=10$, the proposed hybrid storage mechanism can be applied by first determining the basic achievable $\left(\mu=\frac{r}{NK_r},C_T(\mu)\right)$ pairs for $r=4,\dotsc,N$, $K_r=1,\dotsc,r-3$ with $(r-K_r-1)\!\!\mod2=0$. The pairs of $(r,K_r)$ and the corresponding values of $\mu$ are shown in Fig.~\ref{fig2}. The resulting pairs of $(\mu,C_T(\mu))$, before finding the convex hull, are shown in Fig.~\ref{fig3}. Note that each achievable $(\mu,C_T(\mu))$ pair of the hybrid scheme lies on one of the lower bounds characterized in \eqref{lbcoded} and \eqref{lbuncoded}. For instance, $(0.25,10)$ on the red curve in Fig.~\ref{fig3} is on the lower bound that corresponds to $r=5$ in \eqref{lbuncoded} as shown by the pink dotted line. However, since we are interested in finding the lowest possible total costs, from Lemma~\ref{lem1}, the lowest possible total cost is characterized by the boundary of the lower convex hull of all achievable points of the hybrid scheme as shown in Fig.~\ref{fig4}. Note that the set of basic achievable $\left(\mu=\frac{r}{NK_r},C_T(\mu)\right)$ pairs on the lower convex hull boundary corresponds to $\mu$'s with $(r,K_r)$ pairs with $r=N=10$, $r=N-1=9$ and $r=N-2=8$ with corresponding $K_r$'s that satisfy $(r-K_r-1)\!\!\mod2=0$, as marked in Fig.~\ref{fig2}.

\section{A Specific Example}

\begin{figure}
\centering
\begin{subfigure}[b]{0.29\columnwidth}
\includegraphics[width=\columnwidth]{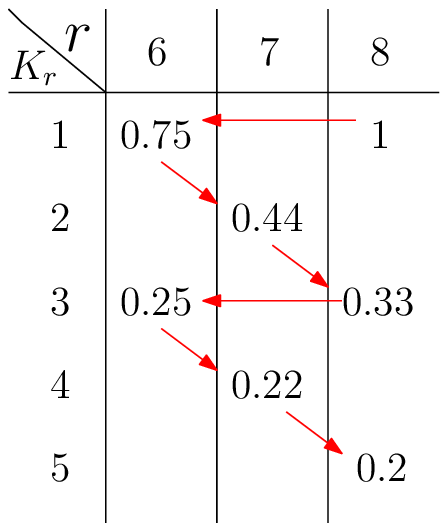}
\caption{$\mu=\frac{r}{NK_r}$ on the boundary.}
\end{subfigure}
\begin{subfigure}[b]{0.68\columnwidth}
\includegraphics[width=\columnwidth]{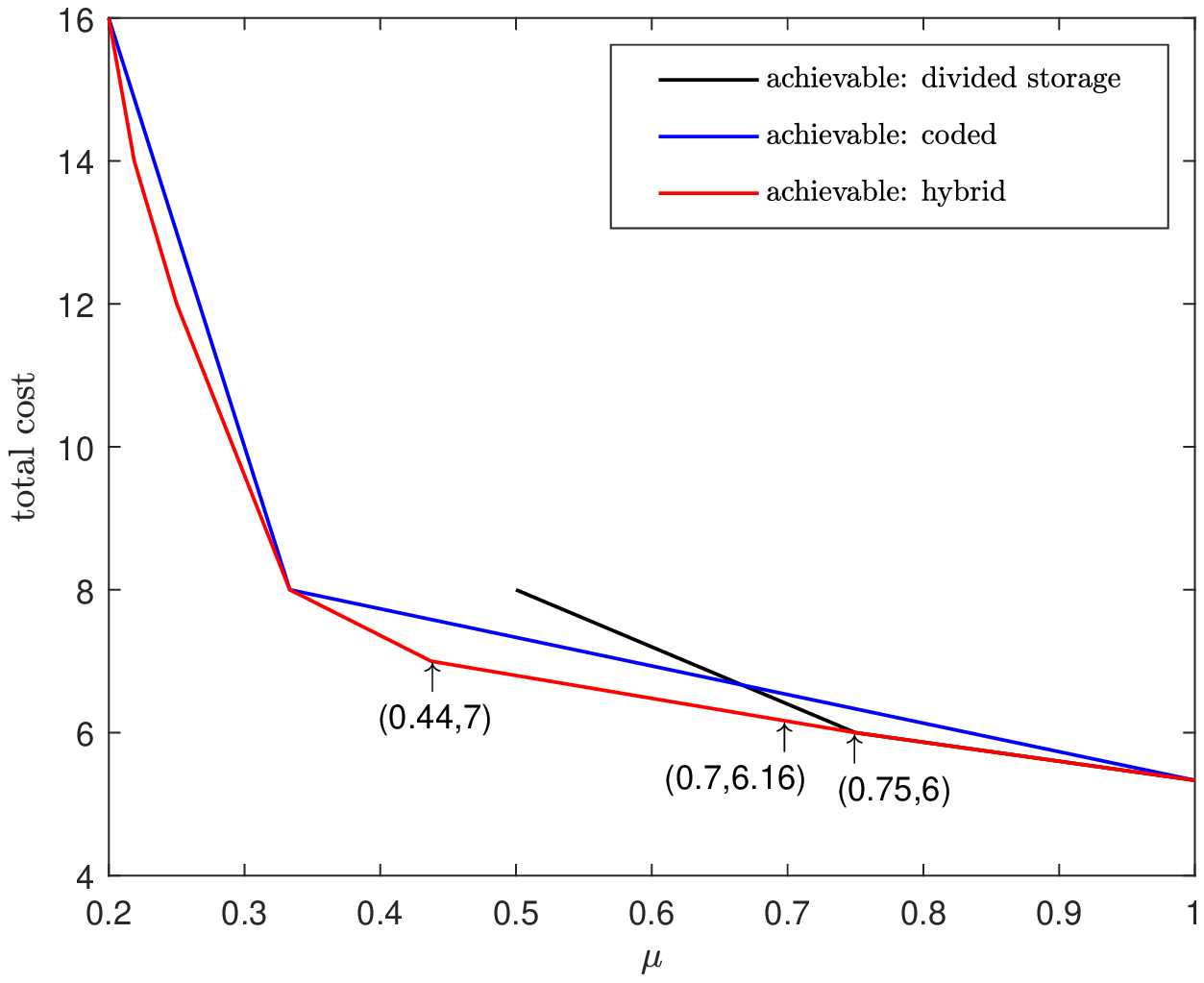}
\caption{Achievable total cost at $\mu=0.7$.}
\end{subfigure}
\caption{Example with $N=8$.}
\label{fig7}
\vspace*{-0.5cm}
\end{figure}

In this section, we provide a complete description on how the PRUW process is carried out in an arbitrary setting with a given $N$ and $\mu$. Consider an example with $N=8$ databases and $\mu=0.7$. The first step is to find the basic achievable $\left(\mu=\frac{r}{NK_r},C_T(\mu)\right)$ pairs of $N=8$ that lie on the lower convex hull boundary. Fig.~\ref{fig7}(a) shows the $(r,K_r)$ pairs and the corresponding $\mu$'s of such pairs.

The required storage constraint $\mu=0.7$ is in between $0.44$ and $0.75$, which correspond to $(r,K_r)$ pairs $(7,2)$ and $(6,1)$, respectively. Therefore, the PRUW scheme for $N=8$, $\mu=0.7$ is obtained by the following steps:
\begin{enumerate}
    \item $\gamma L$ bits of all submodels are stored according to the proposed storage mechanism corresponding to $(r,K_r)=(7,2)$, and the rest of the $(1-\gamma)L$ bits of all submodels are stored according to $(r,K_r)=(6,1)$. Therefore, $\gamma L$ bits of the required submodel are updated using the scheme corresponding to $(r,K_r)=(7,2)$, and the rest of the bits are updated by the scheme corresponding to $(6,1)$. In order to find the value of $\gamma$, we equate the total storage of each database to the given constraint, i.e.,
    \begin{align}
        \gamma ML\times\frac{7}{8}\times\frac{1}{2}+(1-\gamma)ML\times\frac{6}{8}=0.7ML
    \end{align}
    which gives $\gamma=0.16$.
    \item Let $L_1=0.16L$ and $L_2=0.84L$. $L_1$ bits of each submodel is divided into $8$ sections and labeled $1,\dotsc,8$. Sections $n:(n+6)\!\!\mod8$ are allocated to database $n$ for $n\in\{1,\dotsc,N\}$. Each database uses the storage in \eqref{storage} with $K=2$ and $y=x=\frac{r-K_r-1}{2}=2$ to store each subpacket of all sections allocated to it. Then, the PRUW scheme described in Section~\ref{scheme} is applied to read/write to the $L_1$ bits of the required submodel.
     \item The same process is carried out on the rest of the $L_2$ bits with the scheme corresponding to $(6,1)$.
\end{enumerate}

The total costs (normalized) incurred by the two schemes are $C_{T_1}=\frac{4r}{r-K_r-1}=\frac{4\times 7}{7-2-1}=7$ and $C_{T_2}=\frac{4r}{r-K_r-1}=\frac{4\times 6}{6-1-1}=6$, respectively. Therefore, the total cost of $N=8$ and $\mu=0.7$ is $C_T=\frac{\gamma LC_{T_1}+(1-\gamma)LC_{T_2}}{L}=6.16$,
which is shown in Fig.~\ref{fig7}(b).

In both examples with $N=10$ and $N=8$, the boundary of the lower convex hull of the achievable $(\mu,C_T(\mu))$ points was determined by simply connecting the points $(\mu=\frac{r}{NK_r},C_T(\mu))$ with $r=N,N-1,N-2$ with all possible values of $K_r$. In general, we have the following result.

\begin{lemma}\label{lem3}
For any given $N$, let $T_{ach}^{[h]}$ be the piecewise linear curve obtained by connecting the achievable points of the hybrid scheme given by $(\mu=\frac{r}{NK_r},C_T(\mu))$ that correspond to $r=N,N-1,N-2$ with $K_r=1,\dotsc,r-3$ and $(r-K_r-1)\mod2=0$. Let $T_{ach}^{[d]}$ and $T_{ach}^{[c]}$ be the minimum achievable total costs of the divided storage and coded storage schemes, respectively. Then,
\begin{align}
    T_{ach}^{[h]}(\mu)\leq \min\{T_{ach}^{[d]}(\mu),T_{ach}^{[c]}(\mu)\}, \ \ \forall \mu\in[\frac{1}{N-3},1].
\end{align}
\end{lemma}

The proof of Lemma~\ref{lem3} is based on the geometric placement of the points $(\mu=\frac{r}{NK_r},C_T(\mu))$ that correspond to $r=N,N-1,N-2$ with $K_r=1,\dotsc,r-3$ and $(r-K_r-1)\mod2=0$ on the lowest three non-intersecting lower bounds given in \eqref{lbuncoded}, along with the condition in \eqref{range}.

\newpage
\bibliographystyle{unsrt}
\bibliography{references_ISIT}

\begin{thebibliography}{10}

\bibitem{billion}
C.~Niu, F.~Wu, S.~Tang, L.~Hua, R.~Jia, C.~Lv, Z.~Wu, and G.~Chen.
\newblock Billion-scale federated learning on mobile clients: A submodel design
  with tunable privacy.
\newblock In {\em MobiCom}, April 2020.

\bibitem{paper1}
M.~Kim and J.~Lee.
\newblock Information-theoretic privacy in federated submodel learning.
\newblock Available online at arXiv:2008.07656.

\bibitem{secureFSL}
C.~Niu, F.~Wu, S.~Tang, L.~Hua, R.~Jia, C.~Lv, Z.~Wu, and G.~Chen.
\newblock Secure federated submodel learning.
\newblock Available online at arXiv:1911.02254.

\bibitem{rw_jafar}
Z.~Jia and S.~A. Jafar.
\newblock {$X$}-secure {$T$}-private federated submodel learning.
\newblock In {\em IEEE ICC}, June 2021.

\bibitem{ourICC}
S.~Vithana and S.~Ulukus.
\newblock Efficient private federated submodel learning.
\newblock In {\em IEEE ICC}, May 2022.

\bibitem{dropout}
Z.~Jia and S.~A. Jafar.
\newblock ${X}$ -secure ${T}$-private federated submodel learning with elastic
  dropout resilience.
\newblock Available online at arXiv:2010.01059.

\bibitem{aggregation}
C.~Naim, R.~D'Oliveira, and S.~El Rouayheb.
\newblock Private multi-group aggregation.
\newblock {\em IEEE ISIT}, July 2021.

\bibitem{FL1}
H.~B. McMahan, E.~Moore, D.~Ramage, S.~Hampson, and B.~A. y~Arcas.
\newblock Communication efficient learning of deep networks from decentralized
  data.
\newblock {\em AISTATS}, April 2017.

\bibitem{Advances}
P.~Kairouz, H.~B. McMahan, B.~Avent, A.~Bellet, and M.~Bennis \textit{et al.}
\newblock Advances and open problems in federated learning.
\newblock {\em Foundations and Trends in Machine Learning}, 14(1-2):1--210,
  June 2021.

\bibitem{FL2}
Q.~Yang, Y.~Liu, T.~Chen, and Y.~Tong.
\newblock Federated machine learning: Concept and applications.
\newblock {\em ACM Transactions on Intelligent Systems and Technology},
  10(2):1--19, January 2019.

\bibitem{magazine}
T.~Li, A.~K. Sahu, A.~S. Talwalkar, and V.~Smith.
\newblock Federated learning: Challenges, methods, and future directions.
\newblock {\em IEEE Signal Processing Magazine}, 37:50--60, May 2020.

\bibitem{keyboard}
A.~Hard, K.~Rao, R.~Mathews, S.~Ramaswamy, F.~Beaufays, S.~Augenstein,
  H.~Eichner, C.~Kiddon, and D.~Ramage.
\newblock Federated learning for mobile keyboard prediction.
\newblock Available at arXiv: 1811.03604.

\bibitem{recent}
S.~Ulukus, S.~Avestimehr, M.~Gastpar, S.~A. Jafar, R.~Tandon, and C.~Tian.
\newblock Private retrieval, computing and learning: Recent progress and future
  challenges.
\newblock Available online at arxiv:2108.00026.

\bibitem{comprehensive}
M.~Nasr, R.~Shokri, and A.~Houmansadr.
\newblock Comprehensive privacy analysis of deep learning: Passive and active
  white-box inference attacks against centralized and federated learning.
\newblock In {\em {IEEE} Symposium on Security and Privacy}, May 2019.

\bibitem{featureLeakage}
L.~Melis, C.~Song, E.~De Cristofaro, and V.~Shmatikov.
\newblock Exploiting unintended feature leakage in collaborative learning.
\newblock In {\em {IEEE} Symposium on Security and Privacy}, May 2019.

\bibitem{InvertingGradients}
J.~Geiping, H.~Bauermeister, H.~Droge, and M.~Moeller.
\newblock Inverting gradients--how easy is it to break privacy in federated
  learning?
\newblock Available online at arXiv:2003.14053.

\bibitem{client}
R.~C. Geyer, T.~Klein, and M.~Nabi.
\newblock Differentially private federated learning: A client level
  perspective.
\newblock In {\em NeurIPS}, December 2017.

\bibitem{DPFL}
S.~Asoodeh and F.~Calmon.
\newblock Differentially private federated learning: An information-theoretic
  perspective.
\newblock In {\em ICML-FL}, July 2020.

\bibitem{language}
H.~B. McMahan, D.~Ramage, K.~Talwar, and L.~Zhang.
\newblock Learning differentially private recurrent language models.
\newblock In {\em ICLR}, May 2018.

\bibitem{PracticalSecureAgg}
K.~Bonawitz, V.~Ivanov, B.~Kreuter, A.~Marcedone, H.~B. McMahan, S.~Patel,
  D.~Ramage, A.~Segal, and K.~Seth.
\newblock Practical secure aggregation for privacy-preserving machine learning.
\newblock In {\em CCS}, October 2017.

\bibitem{avgDP}
Y.~Li, T.~Chang, and C.~Chi.
\newblock Secure federated averaging algorithm with differential privacy.
\newblock {\em IEEE MLSE}, September 2020.

\bibitem{cpSGD}
N.~Agarwal, A.~Suresh, F.~Yu, S.~Kumar, and H.~B. McMahan.
\newblock cp{SGD}: Communication-efficient and differentially-private
  distributed {SGD}.
\newblock In {\em NeurIPS}, December 2018.

\bibitem{DP}
C.~Dwork and A.~Roth.
\newblock The algorithmic foundations of differential privacy.
\newblock {\em Foundations and Trends in Theoretical Computer Science},
  9(3-4):211--407, August 2014.

\bibitem{OTP}
C.~E. Shannon.
\newblock Communication theory of secrecy systems.
\newblock {\em Bell System Technical Journal}, 28(4):656--715, October 1949.

\bibitem{original}
B.~Chor, E.~Kushilevitz, O.~Goldreich, and M.~Sudan.
\newblock Private information retrieval.
\newblock {\em Journal of the ACM}, 45(6):965--981, November 1998.

\bibitem{PIR}
H.~Sun and S.~A. Jafar.
\newblock The capacity of private information retrieval.
\newblock {\em IEEE Transactions on Information Theory}, 63(7):4075--4088, July
  2017.

\bibitem{coded}
K.~Banawan and S.~Ulukus.
\newblock The capacity of private information retrieval from coded databases.
\newblock {\em IEEE Transactions on Information Theory}, 64(3):1945--1956,
  March 2018.

\bibitem{ravi_storage_constrained}
M.~Attia, D.~Kumar, and R.~Tandon.
\newblock The capacity of private information retrieval from uncoded storage
  constrained databases.
\newblock {\em IEEE Transactions on Information Theory}, 66(11):6617--6634,
  November 2020.

\bibitem{utah}
N.~Woolsey, R.~Chen, and M.~Ji.
\newblock Uncoded placement with linear sub-messages for private information
  retrieval from storage constrained databases.
\newblock {\em IEEE Transactions on Communications}, 68(10):6039--6053, October
  2020.

\bibitem{ITW_paper}
K.~Banawan, B.~Arasli, and S.~Ulukus.
\newblock Improved storage for efficient private information retrieval.
\newblock In {\em IEEE ITW}, August 2019.

\bibitem{Kumar_PIRarbCoded}
S.~Kumar, H.-Y. Lin, E.~Rosnes, and A.~G. i~Amat.
\newblock Achieving maximum distance separable private information retrieval
  capacity with linear codes.
\newblock {\em IEEE Transactions on Information Theory}, 65(7):4243--4273, July
  2019.

\bibitem{YamamotoPIR}
T.~Chan, S.~Ho, and H.~Yamamoto.
\newblock Private information retrieval for coded storage.
\newblock In {\em IEEE ISIT}, June 2015.

\bibitem{VardyConf2015}
A.~Fazeli, A.~Vardy, and E.~Yaakobi.
\newblock Codes for distributed {PIR} with low storage overhead.
\newblock In {\em IEEE ISIT}, June 2015.

\bibitem{MultiroundPIR}
H.~Sun and S.~A. Jafar.
\newblock Multiround private information retrieval: Capacity and storage
  overhead.
\newblock {\em IEEE Transactions on Information Theory}, 64(8):5743--5754,
  August 2018.

\bibitem{heteroPIR}
K.~Banawan, B.~Arasli, Y.-P. Wei, and S.~Ulukus.
\newblock The capacity of private information retrieval from heterogeneous
  uncoded caching databases.
\newblock {\em IEEE Transactions on Information Theory}, 66(6):3407--3416, June
  2020.

\bibitem{decentralized}
Y.-P. Wei, B.~Arasli, K.~Banawan, and S.~Ulukus.
\newblock The capacity of private information retrieval from decentralized
  uncoded caching databases.
\newblock {\em Information}, 10(12):372--389, December 2019.

\bibitem{chaoTian}
C.~Tian, H.~Sun, and J.~Chen.
\newblock Capacity-achieving private information retrieval codes with optimal
  message size and upload cost.
\newblock {\em IEEE Transactions on Information Theory}, 65(11):7613--7627,
  November 2019.

\bibitem{leaky}
I.~Samy, M.~Attia, R.~Tandon, and L.~Lazos.
\newblock Asymmetric leaky private information retrieval.
\newblock {\em IEEE Transactions on Information Theory}, 67(8):5352--5369,
  August 2021.

\bibitem{SPIR}
H.~Sun and S.~A. Jafar.
\newblock The capacity of symmetric private information retrieval.
\newblock {\em IEEE Transactions on Information Theory}, 65(1):329--322,
  January 2019.

\bibitem{PSI}
Z.~Wang, K.~Banawan, and S.~Ulukus.
\newblock Private set intersection: A multi-message symmetric private
  information retrieval perspective.
\newblock Available at arXiv: 1912.13501.

\bibitem{colluding}
H.~Sun and S.~A. Jafar.
\newblock The capacity of robust private information retrieval with colluding
  databases.
\newblock {\em IEEE Transactions on Information Theory}, 64(4):2361--2370,
  April 2018.

\bibitem{CodeColludeByzantinePIR}
R.~Tajeddine, O.~W. Gnilke, D.~Karpuk, R.~Freij-Hollanti, and C.~Hollanti.
\newblock Private information retrieval from coded storage systems with
  colluding, {B}yzantine, and unresponsive servers.
\newblock {\em IEEE Transactions on Information Theory}, 65(6):3898--3906, June
  2019.

\bibitem{sideinfo}
S.~Kadhe, B.~Garcia, A.~Heidarzadeh, S.~El Rouayheb, and A.~Sprintson.
\newblock Private information retrieval with side information.
\newblock {\em IEEE Transactions on Information Theory}, 66(4):2032--2043,
  April 2020.

\bibitem{MMPIR}
K.~Banawan and S.~Ulukus.
\newblock Multi-message private information retrieval: Capacity results and
  near-optimal schemes.
\newblock {\em IEEE Transactions on Information Theory}, 64(10):6842--6862,
  October 2018.

\bibitem{MMPIR_PrivateSideInfo}
M.~J. Siavoshani, S.~P. Shariatpanahi, and M.~A. Maddah-Ali.
\newblock Private information retrieval for a multi-message scenario with
  private side information.
\newblock {\em IEEE Trans. on Commun.}, 69(5):3235--3244, May 2021.

\bibitem{evesdroppers}
Q.~Wang, H.~Sun, and M.~Skoglund.
\newblock The capacity of private information retrieval with eavesdroppers.
\newblock {\em IEEE Transactions on Information Theory}, 65(5):3198--3214, May
  2019.

\bibitem{byzantine}
K.~Banawan and S.~Ulukus.
\newblock The capacity of private information retrieval from {B}yzantine and
  colluding databases.
\newblock {\em IEEE Transactions on Information Theory}, 65(2):1206--1219,
  February 2019.

\bibitem{semanticPIR}
S.~Vithana, K.~Banawan, and S.~Ulukus.
\newblock Semantic private information retrieval.
\newblock {\em IEEE Transactions on Information Theory}, December 2021.

\bibitem{singleDB}
S.~Li and M.~Gastpar.
\newblock Single-server multi-message private information retrieval with side
  information: the general cases.
\newblock In {\em IEEE ISIT}, June 2020.

\bibitem{XSTPIR}
Z.~Jia and S.~A. Jafar.
\newblock {$X$}-secure {$T$}-private information retrieval from {MDS} coded
  storage with {B}yzantine and unresponsive servers.
\newblock {\em IEEE Transactions on Information Theory}, 66(12):7427--7438,
  December 2020.

\bibitem{asympXSTP}
Z.~Jia and S.~Jafar.
\newblock On the asymptotic capacity of ${X}$-secure ${T}$-private information
  retrieval with graph-based replicated storage.
\newblock {\em IEEE Transactions on Information Theory}, 66(10):6280--6296,
  October 2020.

\bibitem{smallfields}
J.~Xu and Z.~Zhang.
\newblock Building capacity-achieving {PIR} schemes with optimal
  sub-packetization over small fields.
\newblock In {\em IEEE ISIT}, June 2018.

\bibitem{SecureStorage}
H.~Yang, W.~Shin, and J.~Lee.
\newblock Private information retrieval for secure distributed storage systems.
\newblock {\em IEEE Transactions on Information Forensics and Security},
  13(12):2953--2964, December 2018.

\end{thebibliography}
\end{document}